\newcommand{\mii}{Mg\,\textsc{ii}}
\newcommand{\civ}{C\,\textsc{iv}}

\documentclass{ws-ijmpd}
\usepackage{hyperref}
\usepackage{url}
\usepackage[usenames]{color}
\usepackage[super,compress]{cite}
\begin{document}

\markboth{Park, de Cruz P\'erez \& Ratra}
{Is $w_0w_a$CDM evidence for DE dynamics partially caused by excess smoothing of Planck data?}

%
\catchline{}{}{}{}{}
%

\title{\boldmath Is the $w_0w_a$CDM cosmological parameterization evidence for dark energy dynamics partially caused by the excess smoothing of Planck CMB anisotropy data?}

\author{Chan-Gyung Park}
\address{Division of Science Education and Institute of Fusion Science, Jeonbuk National University,\\
Jeonju 54896, Republic of Korea\\
park.chan.gyung@gmail.com}

\author{Javier de Cruz P\'erez}
\address{Departamento de F\'isica, Universidad de C\'ordoba, Campus Universitario de Rabanales, Ctra. N-IV km, 396, E-14071, C\'ordoba, Spain\\
jdecruz@uco.es}

\author{Bharat Ratra}
\address{Department of Physics, Kansas State University, 116 Cardwell Hall, Manhattan, KS 66506, USA\\
ratra@ksu.edu}

\maketitle

\begin{history}
\received{Day Month Year}
\revised{Day Month Year}
\end{history}

\begin{abstract}
We study the performance of the spatially-flat dynamical dark energy $w_0w_a$CDM parameterization, with redshift-dependent dark energy fluid equation of state parameter $w(z) = w_0 + w_a z/(1+z)$, with and without a varying CMB lensing consistency parameter $A_L$, against Planck cosmic microwave background (CMB) data (P18 and lensing) and a combination of non-CMB data composed of baryonic acoustic oscillation (BAO) measurements that do not include DESI BAO data, Pantheon+ type Ia supernovae (SNIa) observations, Hubble parameter [$H(z)$] measurements, and growth factor ($f\sigma_8$) data points. From our most restrictive data set, P18+lensing+non-CMB, for the $w_0w_a$CDM+$A_L$ parameterization, we obtain $w_0=-0.879\pm 0.060$, $w_a=-0.39^{+0.26}_{-0.22}$, the asymptotic limit $w(z\to\infty) = w_0+w_a=-1.27^{+0.20}_{-0.17}$, and $A_L=1.078^{+0.036}_{-0.040}$ (all $1\sigma$ errors). This joint analysis of CMB and non-CMB data favors dark energy dynamics over a cosmological constant at $\sim 1\sigma$ and $A_L>1$ at $\sim 2\sigma$, i.e. more smoothing of the Planck CMB anisotropy data than is predicted by the best-fit model. For the $w_0w_a$CDM parameterization with $A_L=1$ the evidence in favor of dark energy dynamics is larger, $\sim 2\sigma$, suggesting that at least part of the evidence for dark energy dynamics comes from the excess smoothing of the Planck CMB anisotropy data. For the $w_0w_a$CDM parameterization with $A_L=1$, there is a difference of $2.8\sigma$ between P18 and non-CMB cosmological parameter constraints and $2.7\sigma$ between P18+lensing and non-CMB constraints. When $A_L$ is allowed to vary these tensions reduced to $1.9\sigma$ and $2.1\sigma$ respectively. Our P18+lensing+non-CMB data compilation {\it positively} favors the $w_0w_a$CDM parameterization without and with a varying $A_L$ parameter over the flat $\Lambda$CDM model, and $w_0w_a$CDM+$A_L$ is also {\it positively} favored over $w_0w_a$CDM.
\end{abstract}


\ccode{PACS numbers: 98.80.-k}


\section{Introduction}
\label{sec:Introduction} 

At cosmological scales the current best description of gravity is general relativity, which provides the framework for the standard model of cosmology ($\Lambda$CDM) \cite{Peebles:1984ge}. This flat $\Lambda$CDM model assumes flat spatial geometry and is characterized by six cosmological parameters. The flat $\Lambda$CDM model cosmological energy budget has contributions from photons, neutrinos, ordinary baryonic matter, cold dark matter (CDM), and a cosmological constant $\Lambda$ that dominates at the present time and so powers the observed accelerated expansion of the universe. This model passes most observational tests but there are some recent measurements that question whether the predictions of the model are correct \cite{Perivolaropoulos:2021jda, Moresco:2022phi, Abdalla:2022yfr, Hu:2023jqc}. 

For instance, the DESI collaboration has recently made public some baryon acoustic oscillation (BAO) measurements \cite{DESI:2024mwx} that may be incompatible with the time-independent $\Lambda$ dark energy component of the flat $\Lambda$CDM model. These authors also study a spatially-flat dynamical dark energy fluid cosmological parameterization, $w_0w_a$CDM, that has a time-evolving dark energy fluid with redshift-dependent equation of state parameter $w(z) = w_0 + w_az/(1+z)$ \cite{Chevallier:2000qy, Linder:2002et} characterized by two degrees of freedom, $w_0$ and $w_a$. When the DESI+CMB+PantheonPlus data set (see Ref.\ \refcite{DESI:2024mwx} for a more detailed description) is analyzed using the $w_0w_a$CDM parameterization they find $w_0 = -0.827 \pm 0.063$ and $w_a = -0.75^{+0.29}_{-0.25}$ with a $\sim 2\sigma$ preference for a time-evolving dark energy over a $\Lambda$. For other discussions of the DESI 2024 results, see Refs.\ \citen{Tada:2024znt, Yin:2024hba, Wang:2024hks, Luongo:2024fww, Cortes:2024lgw, Colgain:2024xqj, Wang:2024rjd, Berghaus:2024kra, Wang:2024rjd, Wang:2024dka, Yang:2024kdo, Park:2024jns, Shlivko:2024llw, Huang:2024qno, DESI:2024aqx, Dinda:2024kjf, Favale:2024sdq, Bhattacharya:2024hep, Ramadan:2024kmn, Mukherjee:2024ryz, Roy:2024kni, Wang:2024hwd, Heckman:2024apk, Gialamas:2024lyw, Notari:2024rti, Liu:2024gfy, Orchard:2024bve, Patel:2024odo, Wang:2024sgo, Li:2024qso, Du:2024pai, Giare:2024gpk, Dinda:2024ktd, Jiang:2024viw, Jiang:2024xnu, Alfano:2024jqn, Ghosh:2024kyd, Reboucas:2024smm, Pang:2024qyh, Wolf:2024eph, RoyChoudhury:2024wri}. In Ref.\ \refcite{Park:2024jns}, we showed, by using a different data set that did not include the DESI BAO measurements, that the $\sim 2\sigma$ preference for a dynamical dark energy component over a cosmological constant did not depend on the DESI measurements, and that our data compilation provides slightly more restrictive constraints, giving $w_0=-0.850\pm 0.059$ and $w_a=-0.59^{+0.26}_{-0.22}$. In Ref.\ \refcite{Park:2024jns} we also showed that the $w_0w_a$CDM parameterization $\sim2\sigma$ preference for dark energy dynamics over a $\Lambda$ also did not depend on Pantheon+ type Ia supernova (SNIa) data. Earlier discussions about the possibility of having a dynamical dark energy component can be found in Refs.\ \citen{Sola:2016hnq, Ooba:2017lng, Ooba:2018dzf, Park:2018fxx,SolaPeracaula:2018wwm, Park:2019emi, Cao:2020jgu, Khadka:2020hvb, Cao:2022wlg, Cao:2022ugh, Dong:2023jtk, VanRaamsdonk:2023ion, deCruzPerez:2024abc, VanRaamsdonk:2024sdp} and references therein. 

It is important to bear in mind that these results are not that statistically significant and also that $w_0w_a$CDM is not a physically consistent cosmological model but just a redshift-dependent parameterization of a dynamical dark energy equation of state. In the simplest physically consistent dynamical dark energy models, the dark energy component is described in terms of an evolving scalar field $\phi$ with a potential energy density $V(\phi)$ \cite{Peebles:1987ek, Ratra:1987rm}. For recent discussions of scalar field dark energy models in the context of the DESI measurements see Refs.\ \citen{Tada:2024znt, Yin:2024hba, Berghaus:2024kra, Yang:2024kdo, Shlivko:2024llw, Bhattacharya:2024hep, Ramadan:2024kmn, Notari:2024rti}. 

In Ref.\ \refcite{Park:2024jns}, within the context of the $w_0w_a$CDM parameterization a difference of about $2.7\sigma$ was found between cosmological parameter constraints obtained with CMB and non-CMB data. One main aim of this work is to determine whether the addition of the variable lensing consistency parameter $A_L$ \cite{Calabreseetal2008} to the dynamical dark energy $w_0w_a$CDM parameterization can help improve its performance when simultaneously fitting different CMB and non-CMB data sets, as is the case in XCDM models \cite{deCruzPerez:2024abc}. Another main aim is to determine whether the flat $w_0w_a$CDM$+A_L$ parameterization better fits these data than does the flat $\Lambda$CDM model and the flat $w_0w_a$CDM parameterization. We find that both of these are true. More importantly, we also find that when $A_L$ is allowed to vary in the $w_0w_a$CDM$+A_L$ parameterization, the evidence for dark energy dynamics over a $\Lambda$ decreases to $\sim 1\sigma$ (compared to the $\sim 2\sigma$ evidence in the $w_0w_a$CDM parameterization case) and that $A_L > 1$ is favored at $\sim 2\sigma$, i.e., that these data prefer more weak lensing of the CMB than is predicted by the best-fit model. These results suggest that at least part of the support for dark energy dynamics in the $w_0w_a$CDM parameterization comes from the excess smoothing of the Planck CMB anisotropy data. Also see Ref.\ \refcite{Chan-GyungPark:2025cri} for similar results in other $w(z)$CDM parameterizations.      

A brief description of the structure of the article follows. In Sec.\ \ref{sec:Data} we provide general details of the different data sets we use to constrain the cosmological parameters and also to test the models under study. A brief summary of the main features of the analysis can be found in Sec.\ \ref{sec:Methods}. In Sec.\ \ref{sec:ResultsandDiscussion} our main results are presented and discussed and finally in Sec.\ \ref{sec:Conclusion} we deliver our conclusions.

\section{Data}
\label{sec:Data}

Here we list the data used in our analyses and the corresponding references, but the details are provided in Sec.\ II of Ref.\ \refcite{deCruzPerez:2024abc}. We note that we account for all known data covariances.

The CMB data sets used in this work are composed of the Planck 2018 TT,TE,EE+lowE (P18) CMB temperature and polarization power spectra \cite{Planck:2018vyg}, which are analyzed alone or in combination with the Planck lensing potential (lensing) power spectrum \cite{planck:2018lbu}. 

The non-CMB data utilized in the analyses here are the same as those denoted non-CMB (new) data that are used in Ref.\ \refcite{deCruzPerez:2024abc} and comprised of 

\begin{itemize}

\item 16 BAO data points, spanning $0.122 \le z \le 2.334$, listed in Table I of Ref.\ \refcite{deCruzPerez:2024abc}. We do not use DESI 2024 BAO data \cite{DESI:2024mwx}.

\item 1590 SNIa data points, a subset of the Pantheon+ compilation \cite{Brout:2022vxf}, where the SNIa at $z < 0.01$ were not used so as to minimize the model-dependency of the peculiar velocity corrections. The range covered by these data is $0.01016 \le z \le 2.26137$.

\item 32 Hubble parameter [$H(z)$]  measurements, spanning $0.070 \le z \le 1.965$, which are listed in Table 1 of Ref.\ \refcite{Cao:2023eja} and also in Table II of Ref.\ \refcite{deCruzPerez:2024abc}. 

\item 9 growth rate ($f\sigma_8$) data points, not obtained from BAO analyses, covering $0.013 \le z \le 1.36$. The complete list is provided in Table III of Ref.\  \refcite{deCruzPerez:2024abc}.

\end{itemize}

We use five individual and combined data sets to constrain the flat  $\Lambda$CDM model and the flat dynamical dark energy $w_0w_a$CDM and $w_0w_a$CDM+$A_L$ parameterizations, namely: P18 data, P18+lensing data, non-CMB data, P18+non-CMB data, and P18+lensing+non-CMB data.

\section{Methods}
\label{sec:Methods} 

Here we present a brief summary of the methods used in our study. A fuller discussion can be found in Sec.\ III of Ref.\ \refcite{deCruzPerez:2024abc}. 

In order to pin down the values of the cosmological parameters that better describe these observational data we use the \texttt{CAMB}/\texttt{COSMOMC} program (October 2018 version) \cite{Challinor:1998xk,Lewis:1999bs,Lewis:2002ah}. While \texttt{CAMB} computes the evolution of cosmological model spatial inhomogeneities and makes theoretical predictions, that in turn depend on the cosmological parameters that characterize the different models under study, \texttt{COSMOMC} compares these predictions to observational data using the Markov chain Monte Carlo (MCMC) method to determine the posterior probability distributions of the involved parameters. The MCMC chains are considered to have converged when the Gelman and Rubin $R$ statistic satifies $R-1 < 0.01$.  Once the converged chains are obtained, we utilize the \texttt{GetDist} code \cite{Lewis:2019xzd} to extract the average values, confidence intervals, and likelihood distributions of the cosmological model parameters.

The six flat $\Lambda$CDM model primary cosmological parameters we have chosen to use are the current value of the physical baryonic matter density parameter $\Omega_b h^2$ ($h$ is the Hubble constant in units of 100 km s$^{-1}$ Mpc$^{-1}$), the current value of the physical cold dark matter density parameter $\Omega_c h^2$, the angular size of the sound horizon evaluated at recombination $\theta_{\text{MC}}$, the reionization optical depth $\tau$, the primordial scalar-type perturbation power spectral index $n_s$, and the power spectrum amplitude $A_s$. We use flat priors for these parameters, non-zero over: $0.005 \le \Omega_b h^2 \le 0.1$, $0.001 \le \Omega_c h^2 \le 0.99$, $0.5 \le 100\theta_\textrm{MC} \le 10$, $0.01 \le \tau \le 0.8$, $0.8 \le n_s \le 1.2$, and $1.61 \le \ln(10^{10} A_s) \le 3.91$. 

In the dynamical dark energy $w_0w_a$CDM and $w_0w_a$CDM+$A_L$ parameterizations, dark energy is taken to be a perfect fluid characterized by a time-evolving equation of state parameter $w(z) = w_0 + w_a z/(1+z)$ \cite{Chevallier:2000qy, Linder:2002et}. For the lensing consistency parameter $A_L$ we use a flat prior non-zero over $0 \le A_L \le 10$. For the additional dark energy equation of state parameters we adopt flat priors non-zero over $-3.0 \le w_0 \le 0.2$ and $-3 < w_a < 2$. Due to the inability of non-CMB data to constrain the values of the $\tau$ and $n_s$ parameters, in the corresponding analyses we fix their values to those obtained from P18 data and constrain only the other parameters. In addition, we also present constraints on three derived parameters, namely the Hubble constant $H_0$, the current value of the non-relativistic matter density parameter $\Omega_m$, and the amplitude of matter fluctuations $\sigma_8$, which are obtained from the values of the primary parameters of the cosmological model. Also, we show the value of the sum of dark energy equation of state parameters $w_0+w_a$, to which $w(z)$ asymptotes at high $z$. 

The primordial scalar-type energy density perturbation power spectrum considered for the flat $\Lambda$CDM model and the flat $w_0 w_a$CDM and $w_0 w_a$CDM+$A_L$ parameterizations is 
\begin{equation}
    P_\delta (k) = A_s \left( \frac{k}{k_0} \right)^{n_s},
\label{eq:powden-flat}
\end{equation}
where $k$ is wavenumber and $n_s$ and $A_s$ are the spectral index and the amplitude of the spectrum at pivot scale $k_0=0.05~\textrm{Mpc}^{-1}$. This power spectrum is generated by quantum fluctuations during an early epoch of power-law inflation in a spatially-flat inflation model powered by a scalar field inflaton potential energy density that is an exponential function of the inflaton \cite{Lucchin:1984yf, Ratra:1989uv, Ratra:1989uz}.

To properly quantify how relatively well each model fits the different combinations of data sets under study, we apply two commonly used model selection criteria: the Akaike information criterion (AIC) and the deviance information criterion (DIC). The AIC is defined as $\mathrm{AIC}=\chi_{\textrm{min}}^2+2k$, where $\chi_{\textrm{min}}^2$ the minimum value of $\chi^2$ for the best-fit cosmological parameters and $k$ is the number of independent cosmological parameters. It penalizes model complexity to prevent overfitting, with lower AIC values indicating a better balance between goodness of fit and simplicity. The DIC, on the other hand, is given by $\mathrm{DIC}=\chi^2 (\hat{\theta})+2p_D$, where $p_D=\overline{\chi^2}-\chi^2(\hat{\theta})$ and the term $2 p_D$ penalizes the
goodness-of-fit for increasing the number of model parameters. Here $\overline{\chi^2}$ denotes the average of $\chi^2$'s estimated from the MCMC chains and $\chi^2(\hat{\theta})$ is the value of $\chi^2$ at the best-fit cosmological parameters $\hat{\theta}$.
In our analysis, we evaluate the differences in the Akaike information criterion ($\Delta$AIC) and the deviance information criterion ($\Delta$DIC) between the information criterion (IC) values for the flat dynamical dark energy $w_0 w_a$CDM parameterizations, with and without an $A_L$-varying parameter, and the flat $\Lambda$CDM model. For a more detailed description of these statistical estimators, see Sec.\ III of Ref.\ \refcite{deCruzPerez:2024abc} and references therein. According to Jeffreys' scale, when $-2 \leq \Delta\textrm{IC}<0$ there is {\it weak} evidence in favor of the model under study. For $-6 \leq \Delta\textrm{IC} < -2$ there is {\it positive} evidence, for $-10\leq\Delta\textrm{IC} < -6$ there is {\it strong} evidence, and when $\Delta\textrm{IC} < -10$ we can claim {\it very strong} evidence in favor of the model under study relative to the tilted flat $\Lambda$CDM model. Conversely, if $\Delta\textrm{IC}$ values are positive, the $\Lambda$CDM model is favored over the model under study.


We are interested in quantitatively comparing how consistent the cosmological parameter constraints are, within a given model, when obtained from two different data sets. To do this, we use two different statistical estimators.
The first is $\log_{10}\mathcal{I}$, which measures the consistency between data sets based on DIC values. Here, $\mathcal{I}$ is defined as $\mathcal{I} (D_1, D_2)=\exp [ -\mathcal{G} (D_1, D_2) / 2 ]$, where $\mathcal{G}(D_1, D_2)=\textrm{DIC}(D_{12})-\textrm{DIC}(D_1)-\textrm{DIC}(D_2)$ and $D_1$ and $D_2$ are the datasets being compared; $\textrm{DIC}(D_1)$ and $\textrm{DIC}(D_2)$ are the DIC values estimated from MCMC chains when each dataset is used independently to constrain the model parameters. $\textrm{DIC}(D_{12})$ is the DIC value obtained when both datasets are combined to constrain the cosmological parameters of the model (see Ref.\ \refcite{Joudaki:2016mvz} and Sec.\ III of Ref.\ \refcite{deCruzPerez:2024abc}).
Positive values  ($\log_{10}\mathcal{I}>0$) indicate that the two data sets are consistent, whereas negative values ($\log_{10}\mathcal{I}<0$) indicate inconsistency. According to Jeffreys' scale, the degree of concordance or discordance between two data sets is classified as {\it substantial} if $\lvert \log_{10}\mathcal{I} \rvert >0.5$, {\it strong} if $\lvert \log_{10}\mathcal{I} \rvert >1$, and {\it decisive} if $\lvert \log_{10}\mathcal{I} \rvert >2$ \cite{Joudaki:2016mvz}. The second estimator we consider is the tension probability $p$ and the related Gaussian approximation "sigma value" $\sigma$ (see Refs.\ \citen{Handley:2019pqx, Handley:2019wlz, Handley:2019tkm} and also Sec.\ III of Ref.\ \refcite{deCruzPerez:2024abc}). Approximately, a value of $p=0.05$ corresponds to $2\sigma$ and $p=0.003$ corresponds to a 3$\sigma$ Gaussian standard deviation.


\begin{table}[htbp]
\tbl{Mean and 68\% (or 95\% indicated between parentheses when the value is provided) confidence limits of flat $w_0 w_a$CDM model parameters from non-CMB, P18, P18+lensing, P18+non-CMB, and P18+lensing+non-CMB data. $H_0$ has units of km s$^{-1}$ Mpc$^{-1}$. We also include the values of $\chi^2_{\text{min}}$, DIC, and AIC and the differences with respect to the $\Lambda$CDM model, denoted by $\Delta\chi^2_{\text{min}}$, $\Delta$DIC, and $\Delta$AIC, respectively.}
{\begin{tabular}{@{}cccccc@{}} \toprule
Parameter   &  Non-CMB    & P18   &  P18+lensing   &  P18+non-CMB     & P18+lensing+non-CMB   \\
\colrule
  $\Omega_b h^2$                & $0.0315 \pm 0.0043$          & $0.02240 \pm 0.00015$       & $0.02243 \pm 0.00015$      &  $0.02245 \pm 0.00014$  &  $0.02244 \pm 0.00014$  \\
  $\Omega_c h^2$                & $0.0990^{+0.0061}_{-0.011}$  & $0.1199 \pm 0.0014$         & $0.1192 \pm 0.0012$        &  $0.1190 \pm 0.0011$    &  $0.1191 \pm 0.0010$  \\
  $100\theta_\textrm{MC}$       & $1.0218^{+0.0087}_{-0.011}$  & $1.04094 \pm 0.00031$       & $1.04101 \pm 0.00031$      &  $1.04101 \pm 0.00030$  &  $1.04100 \pm 0.00029$  \\
  $\tau$                        & $0.0540$                     & $0.0540 \pm 0.0079$         & $0.0523 \pm 0.0074$        &  $0.0529 \pm 0.0077$    &  $0.0534 \pm 0.0072$    \\
  $n_s$                         & $0.9654$                     & $0.9654 \pm 0.0043$         & $0.9669 \pm 0.0041$        &  $0.9672 \pm 0.0040$    &  $0.9670 \pm 0.0039$    \\
  $\ln(10^{10} A_s)$            & $3.60\pm 0.24$ ($>3.13$)     & $3.043 \pm 0.016$           & $3.038 \pm 0.014$          &  $3.039 \pm 0.016$      &  $3.040 \pm 0.014$      \\
  $w_0$                         & $-0.876 \pm 0.055$           & $-1.25^{+0.43}_{-0.56}$     & $-1.24^{+0.44}_{-0.56}$    &  $-0.853 \pm 0.061$     &  $-0.850 \pm 0.059$     \\  
  $w_a$                         & $0.10^{+0.32}_{-0.20}$       & $-1.3 \pm 1.2$ ($< 1.13$)   & $-1.2 \pm 1.3$ ($<1.19$)   &  $-0.57^{+0.27}_{-0.23}$&  $-0.59^{+0.26}_{-0.22}$  \\  
\colrule
  $w_0+w_a$                     & $-0.78^{+0.28}_{-0.15}$      & $-2.50^{+0.74}_{-1.2}$      & $-2.40^{+0.80}_{-1.2}$     &  $-1.42^{+0.21}_{-0.18}$&  $-1.44^{+0.20}_{-0.17}$  \\
  $H_0$                         & $69.8\pm 2.4$                & $84 \pm 11$ ($> 64.5$)      & $84 \pm 11$ ($>64.7$)      &  $67.81 \pm 0.64$       &  $67.80 \pm 0.64$       \\
  $\Omega_m$                    & $0.2692^{+0.0086}_{-0.015}$  & $0.213^{+0.016}_{-0.070}$   & $0.213^{+0.017}_{-0.071}$  &  $0.3092 \pm 0.0063$    &  $0.3094 \pm 0.0063$    \\
  $\sigma_8$                    & $0.823^{+0.031}_{-0.027}$    & $0.955^{+0.11}_{-0.050}$    & $0.945^{+0.11}_{-0.048}$   &  $0.810 \pm 0.011$      &  $0.8108 \pm 0.0091$    \\
\colrule
  $\chi_{\textrm{min}}^2$       & $1457.16$                    & $2761.18$                   & $2770.39$                  & $4234.18$               &  $4243.01$              \\
  $\Delta\chi_{\textrm{min}}^2$ & $-12.77$                     & $-4.62$                     & $-4.32$                    & $-6.06$                 &  $-6.25$              \\
  $\textrm{DIC}$                & $1470.93$                    & $2815.45$                   & $2824.19$                  & $4290.48$               &  $4298.75$              \\
  $\Delta\textrm{DIC}$          & $-7.18$                      & $-2.48$                     & $-2.26$                    & $-1.85$                 &  $-2.45$                \\
  $\textrm{AIC}$                & $1469.16$                    & $2819.18$                   & $2828.39$                  & $4292.18$               &  $4301.01$              \\
  $\Delta\textrm{AIC}$          & $-8.77$                      & $-0.62$                     & $-0.32$                    & $-2.05$                 &  $-2.25$                \\
\botrule
\end{tabular}\label{tab:results_flat_w0waCDM}}
\end{table}

\begin{table}[htbp]
\tbl{Mean and 68\% (or 95\%) confidence limits of flat $w_0 w_a$CDM$+A_L$ model parameters from non-CMB, P18, P18+lensing, P18+non-CMB, and P18+lensing+non-CMB data. $H_0$ has units of km s$^{-1}$ Mpc$^{-1}$.}
{\begin{tabular}{@{}cccccc@{}} \toprule
 Parameter   &  Non-CMB    & P18   &  P18+lensing   &  P18+non-CMB     & P18+lensing+non-CMB     \\
\colrule
  $\Omega_b h^2$                & $0.0315 \pm 0.0043$           & $0.02259 \pm 0.00017$      & $0.02250 \pm 0.00017$      &  $0.02264 \pm 0.00015$     &  $0.02256 \pm 0.00015$  \\
  $\Omega_c h^2$                & $0.0990^{+0.0061}_{-0.011}$  & $0.1181 \pm 0.0015$         & $0.1185 \pm 0.0015$        &  $0.1175 \pm 0.0012$       &  $0.1177 \pm 0.0012$  \\
  $100\theta_\textrm{MC}$       & $1.0218^{+0.0087}_{-0.011}$  & $1.04113 \pm 0.00033$       & $1.04108 \pm 0.00033$      &  $1.04120 \pm 0.00031$     &  $1.04115 \pm 0.00030$  \\
  $\tau$                        & $0.0540$                     & $0.0497^{+0.0085}_{-0.0074}$& $0.0495 \pm 0.0085$        &  $0.0485^{+0.0085}_{-0.0073}$ &$0.0482^{+0.0086}_{-0.0074}$    \\
  $n_s$                         & $0.9654$                     & $0.9706 \pm 0.0049$         & $0.9690 \pm 0.0049$        &  $0.9723 \pm 0.0043$       &  $0.9711 \pm 0.0043$    \\
  $\ln(10^{10} A_s)$            & $3.60\pm 0.24$ ($>3.13$)     & $3.030^{+0.018}_{-0.016}$   & $3.030^{+0.018}_{-0.016}$  &  $3.026^{+0.017}_{-0.015}$ &  $3.025^{+0.018}_{-0.015}$      \\
  $A_L$                         & $\ldots$                     & $1.161^{+0.063}_{-0.086}$   & $1.046^{+0.038}_{-0.057}$  &  $1.191 \pm 0.064$         &  $1.078^{+0.036}_{-0.040}$     \\ 
  $w_0$                         & $-0.876 \pm 0.055$           & $-1.07^{+0.54}_{-0.69}$     & $-1.14^{+0.48}_{-0.68}$    &  $-0.880 \pm 0.059$        &  $-0.879 \pm 0.060$     \\
  $w_a$                         & $0.10^{+0.32}_{-0.20}$       & $-0.8 \pm 1.3$ ($< 1.42$)   & $-0.9 \pm 1.3$ ($<1.44$)   &  $-0.37 \pm 0.24$          &  $-0.39^{+0.26}_{-0.22}$     \\
\colrule
  $w_0+w_a$                     & $-0.78^{+0.28}_{-0.15}$      & $-1.91^{+1.4}_{-0.90}$      & $-2.0^{+1.2}_{-1.1}$       &  $-1.25^{+0.20}_{-0.17}$   &  $-1.27^{+0.20}_{-0.17}$   \\
  $H_0$                         & $69.8\pm 2.4$                & $77^{+20}_{-9}$ ($>55.8$)   & $79 \pm 13$ ($>57.6$)      &  $67.87 \pm 0.64$          &  $67.84 \pm 0.64$       \\
  $\Omega_m$                    & $0.2692^{+0.0086}_{-0.015}$  & $0.259^{+0.033}_{-0.12}$    & $0.245^{+0.025}_{-0.10}$   &  $0.3058 \pm 0.0063$       &  $0.3062 \pm 0.0064$    \\
  $\sigma_8$                    & $0.823^{+0.031}_{-0.027}$    & $0.875^{+0.16}_{-0.082}$    & $0.895^{+0.15}_{-0.070}$   &  $0.793 \pm 0.012$         &  $0.794 \pm 0.012$    \\
\colrule
  $\chi_{\textrm{min}}^2$       & $1457.16$                    & $2755.98$                   & $2770.22$                  & $4222.87$                  &  $4238.26$             \\
  $\Delta\chi_{\textrm{min}}^2$ & $-12.77$                     & $-9.82$                     & $-4.49$                    & $-17.37$                    &  $-11.00$              \\
  $\textrm{DIC}$                & $1470.93$                    & $2812.60$                   & $2825.75$                  & $4283.17$                  &  $4296.83$             \\
  $\Delta\textrm{DIC}$          & $-7.18$                      & $-5.33$                     & $-0.70$                    & $-9.16$                    &  $-4.37$              \\
  $\textrm{AIC}$                & $1469.16$                    & $2815.98$                   & $2830.22$                  & $4282.87$                  &  $4298.26$           \\
  $\Delta\textrm{AIC}$          & $-8.77$                      & $-3.82$                     & $+1.51$                    & $-11.37$                    &  $-5.00$              \\
\botrule
\end{tabular}\label{tab:results_flat_w0waCDM_AL}}
\end{table}

\section{Results and Discussion}
\label{sec:ResultsandDiscussion}

\begin{figure*}[htbp]
\centering
\mbox{\includegraphics[width=127mm]{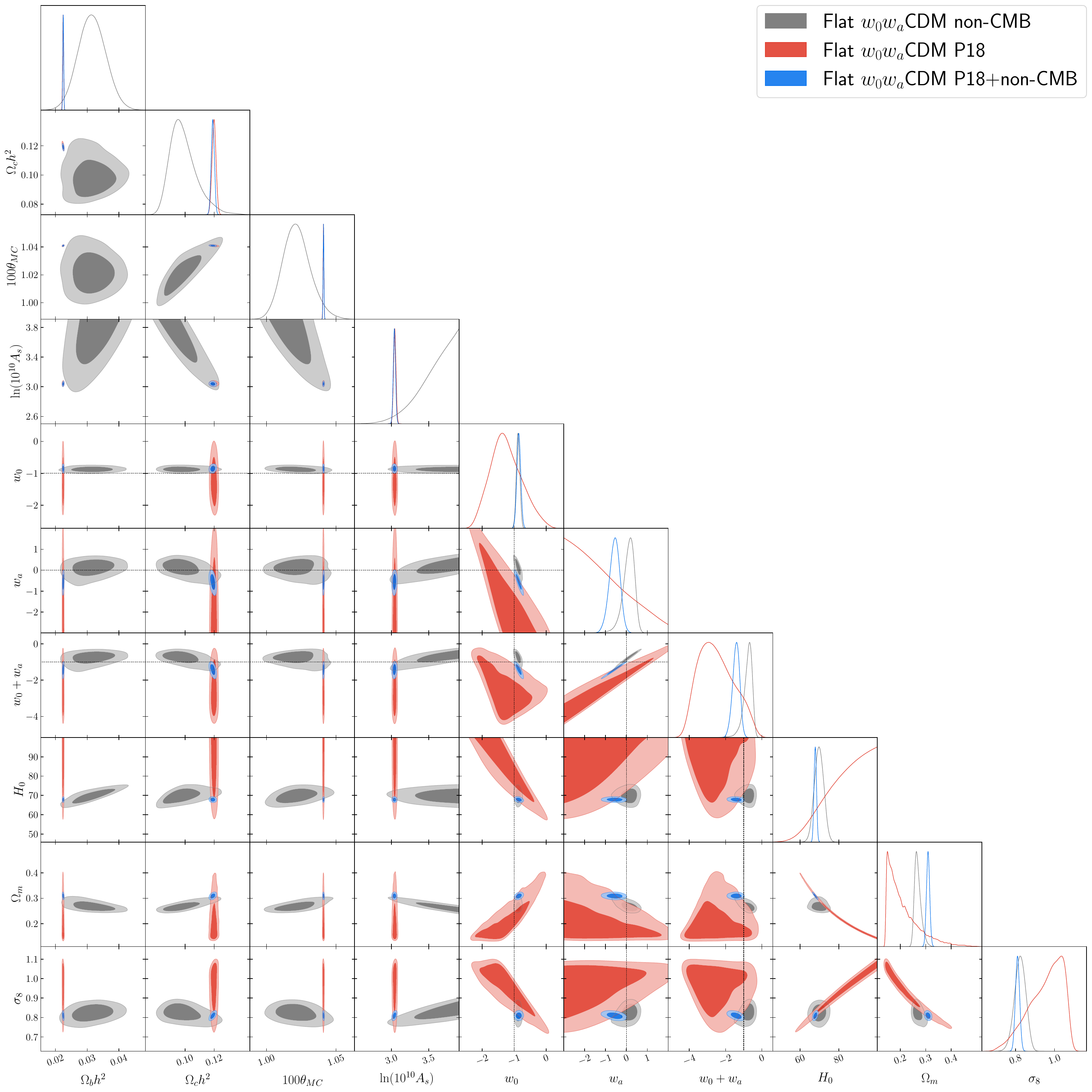}}
        \caption{One-dimensional likelihoods and 1$\sigma$ and $2\sigma$ likelihood confidence contours of flat $w_0 w_a$CDM model parameters favored by non-CMB, P18, and P18+non-CMB data sets. We do not show $\tau$ and $n_s$, which are fixed in the non-CMB data analysis. The horizontal or vertical dotted lines representing $w_0=-1$, $w_a=0$, $w_0+w_a=-1$ correspond to the values in the standard $\Lambda$CDM model.
}
\label{fig:flat_w0waCDM_P18_nonCMB23v2}
\end{figure*}

\begin{figure*}[htbp]
\centering
\mbox{\includegraphics[width=127mm]{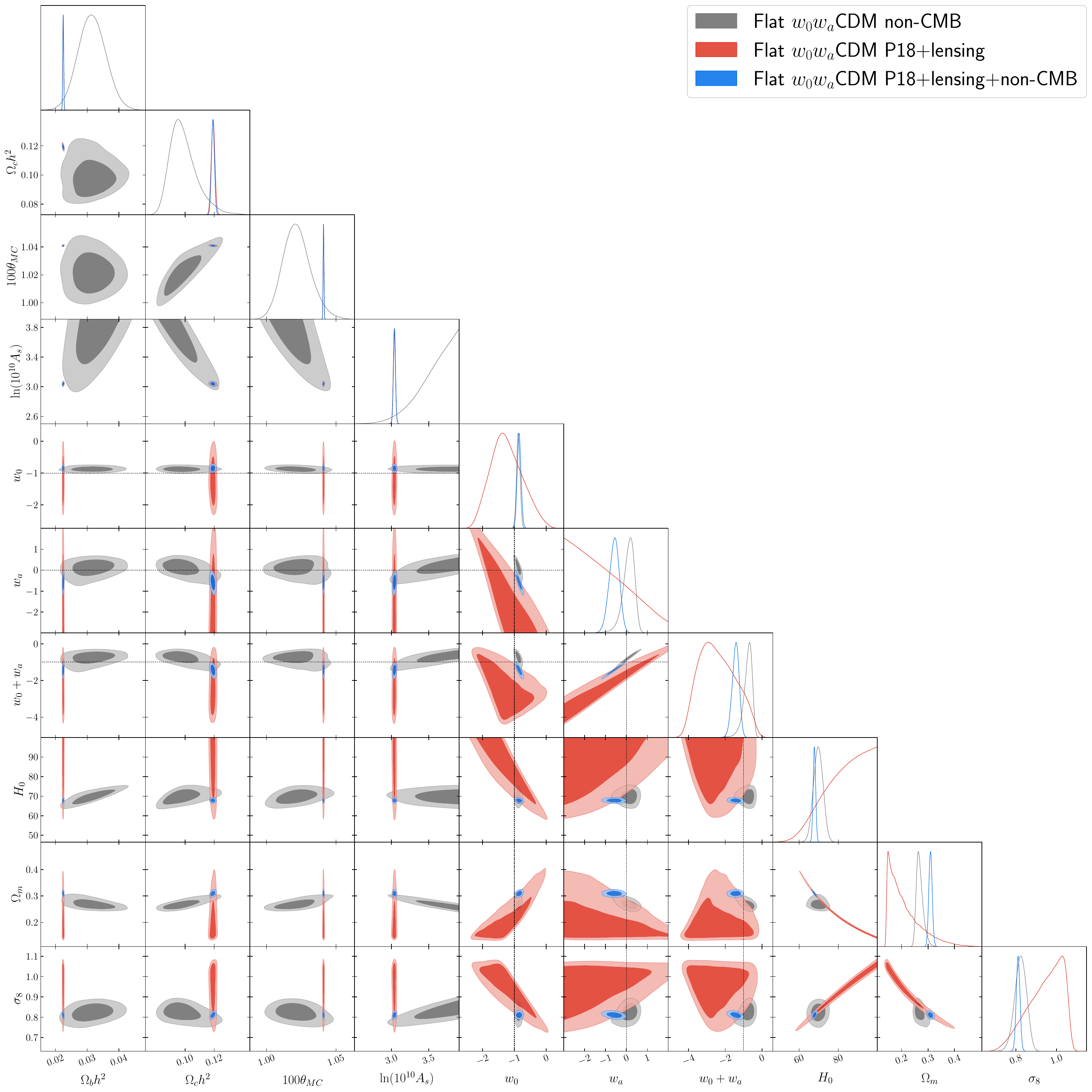}}
\caption{One-dimensional likelihoods and 1$\sigma$ and $2\sigma$ likelihood confidence contours of flat $w_0 w_a$CDM model parameters favored by non-CMB, P18+lensing, P18+lensing+non-CMB data sets. We do not show $\tau$ and $n_s$, which are fixed in the non-CMB data analysis. The horizontal or vertical dotted lines representing $w_0=-1$, $w_a=0$, $w_0+w_a=-1$ correspond to the values in the standard $\Lambda$CDM model.
}
\label{fig:flat_w0waCDM_P18_lensing_nonCMB23v2}
\end{figure*}


\begin{figure*}[htbp]
\centering
\mbox{\includegraphics[width=127mm]{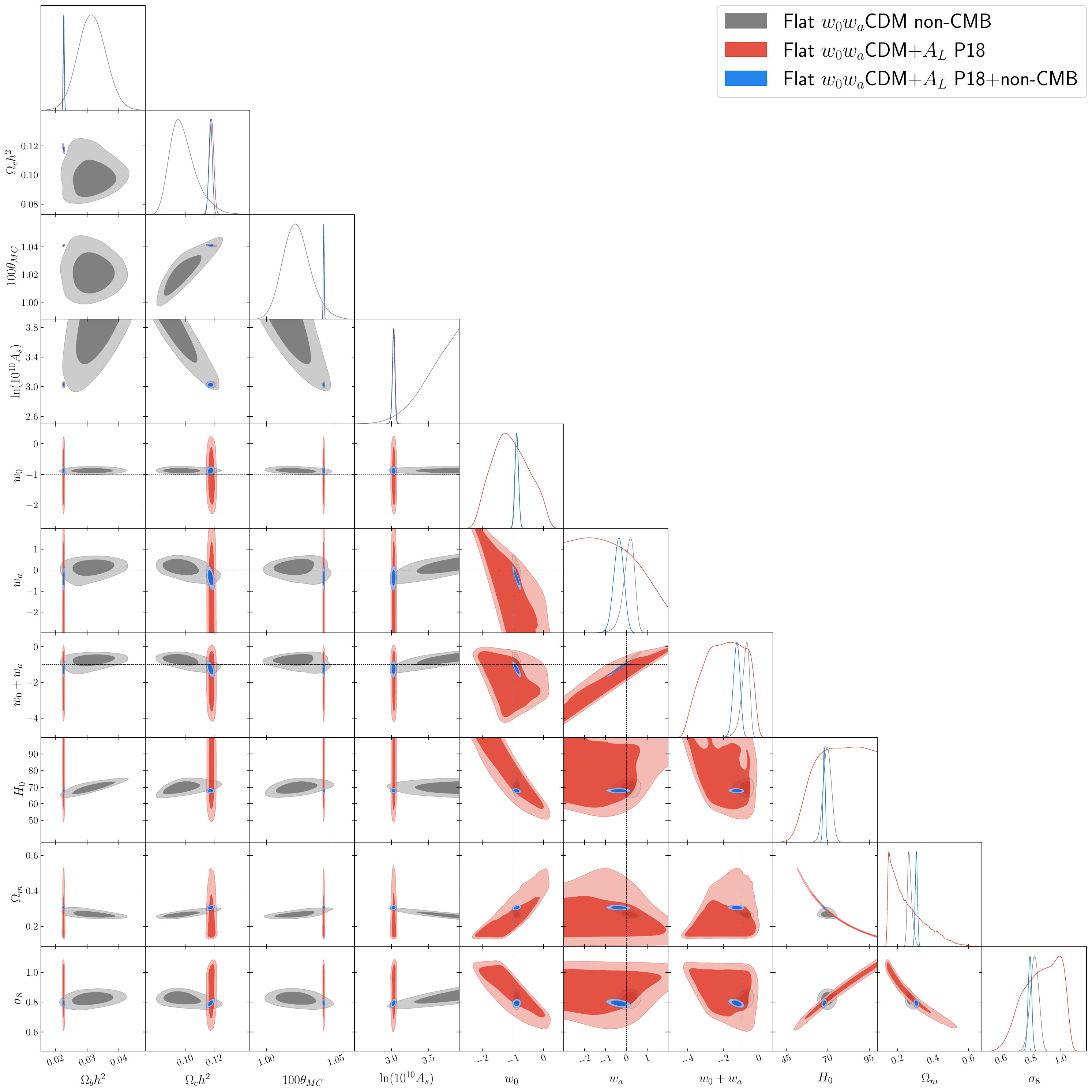}}
        \caption{One-dimensional likelihoods and 1$\sigma$ and $2\sigma$ likelihood confidence contours of flat $w_0 w_a$CDM$+A_L$ model parameters favored by non-CMB, P18, and P18+non-CMB data sets. We do not show $\tau$ and $n_s$, which are fixed in the non-CMB data analysis. The horizontal or vertical dotted lines representing $w_0=-1$, $w_a=0$, $w_0+w_a=-1$ correspond to the values in the standard $\Lambda$CDM model.
}
\label{fig:flat_w0waCDM_Alens_P18_nonCMB23v2}
\end{figure*}

\begin{figure*}[htbp]
\centering
\mbox{\includegraphics[width=127mm]{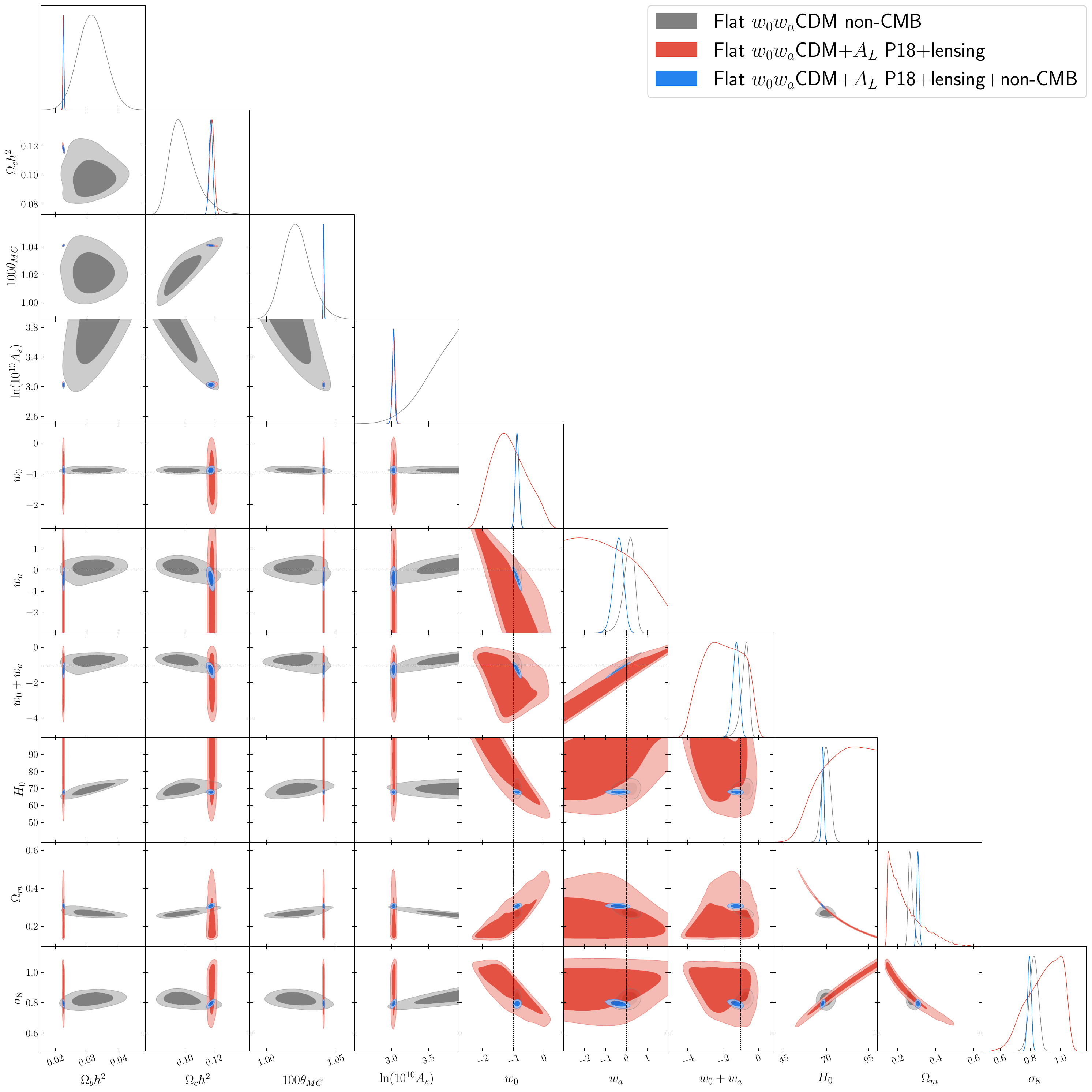}}
\caption{One-dimensional likelihoods and 1$\sigma$ and $2\sigma$ likelihood confidence contours of flat $w_0 w_a$CDM$+A_L$ model parameters favored by non-CMB, P18+lensing, P18+lensing+non-CMB data sets. We do not show $\tau$ and $n_s$, which are fixed in the non-CMB data analysis. The horizontal or vertical dotted lines representing $w_0=-1$, $w_a=0$, $w_0+w_a=-1$ correspond to the values in the standard $\Lambda$CDM model.
}
\label{fig:flat_w0waCDM_Alens_P18_lensing_nonCMB23v2}
\end{figure*}

\begin{figure*}[htbp]
\centering
\mbox{\includegraphics[width=62mm]{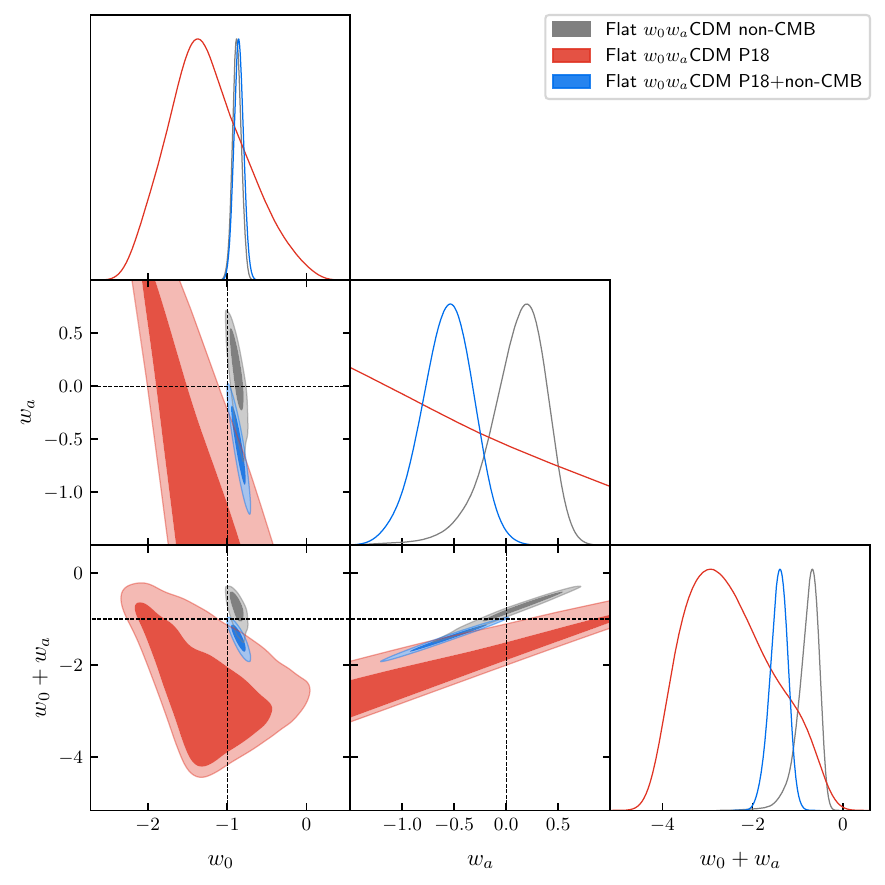}}
\mbox{\includegraphics[width=62mm]{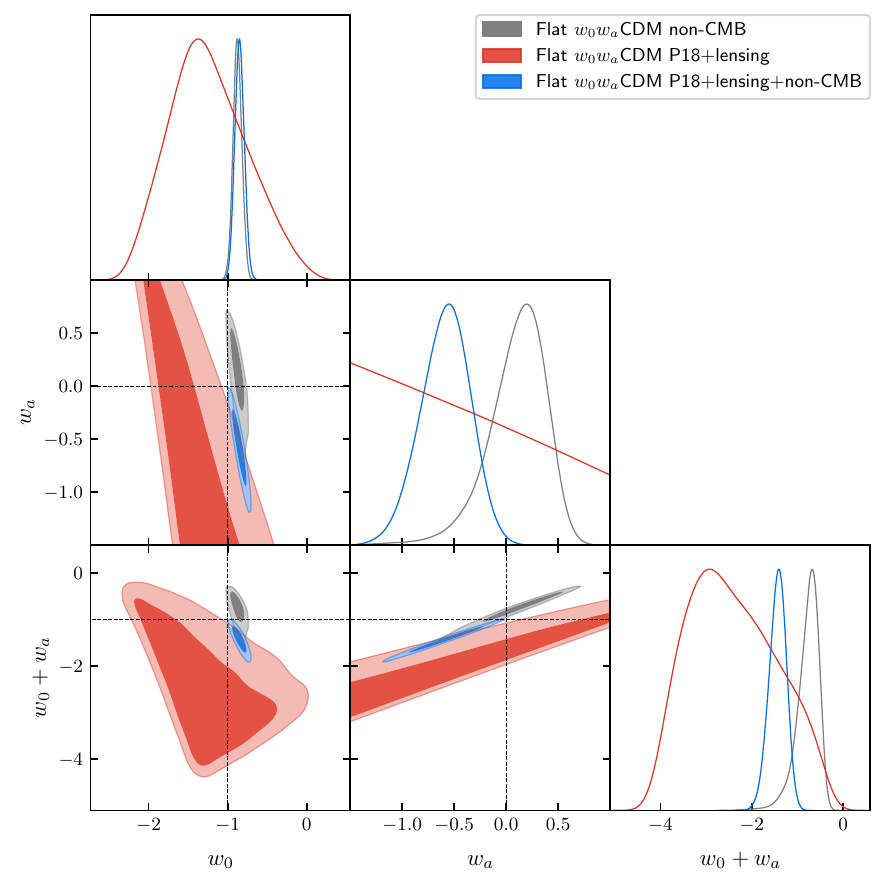}}
        \caption{One-dimensional likelihoods and 1$\sigma$ and $2\sigma$ likelihood confidence contours of $w_0$, $w_a$, and $w_0+w_a$ parameters in the flat $w_0 w_a$CDM model favored by (left) non-CMB, P18, and P18+non-CMB data sets, and (right) non-CMB, P18+lensing, and P18+lensing+non-CMB data sets. The horizontal or vertical dotted lines representing $w_0=-1$, $w_a=0$, $w_0+w_a=-1$ correspond to the values in the standard $\Lambda$CDM model.
}
\label{fig:flat_w0waCDM_P18_nonCMB23v2_w0wa}
\end{figure*}
\begin{figure*}[htbp]
\centering
\mbox{\includegraphics[width=62mm]{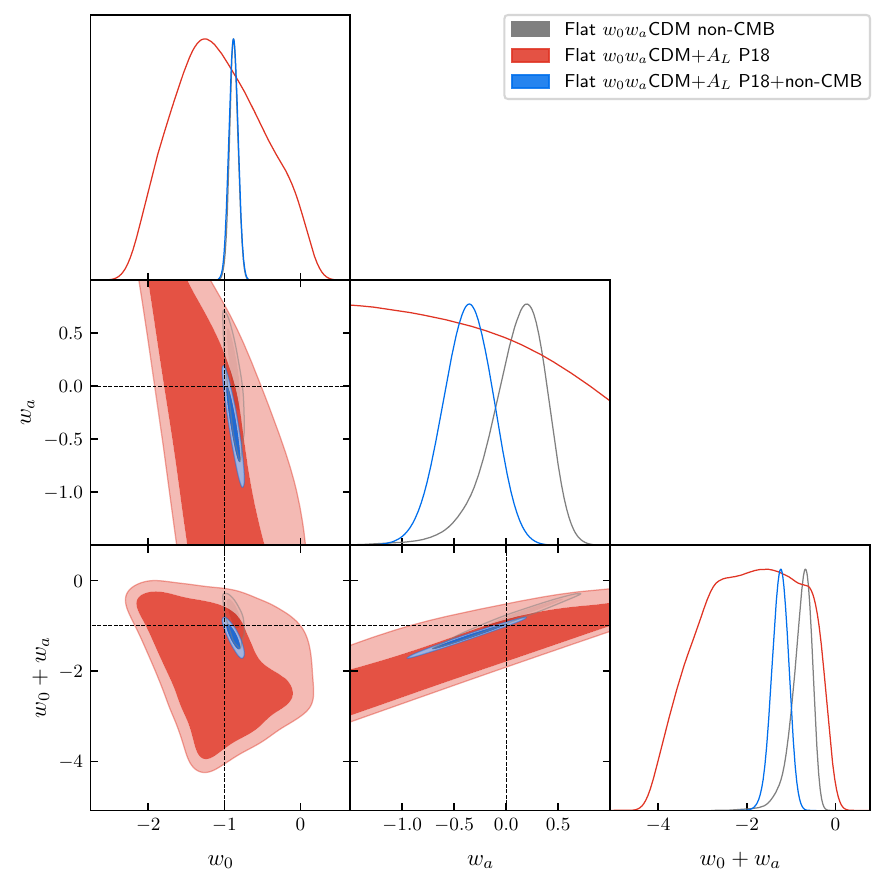}}
\mbox{\includegraphics[width=62mm]{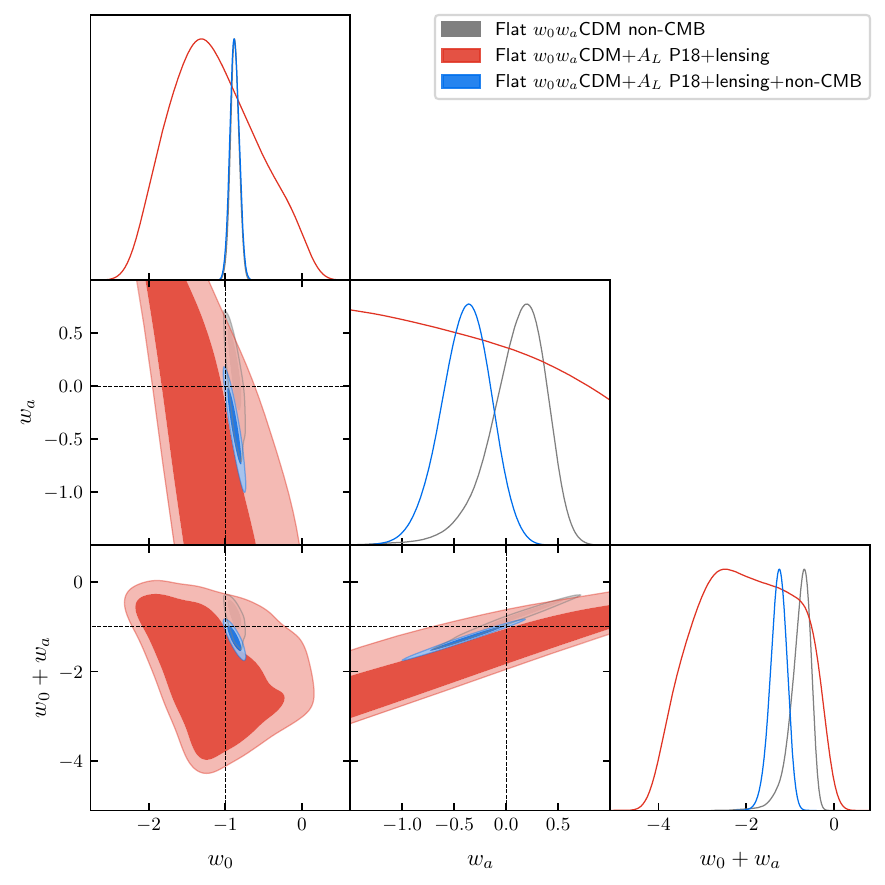}}
        \caption{One-dimensional likelihoods and 1$\sigma$ and $2\sigma$ likelihood confidence contours of $w_0$, $w_a$, and $w_0+w_a$ parameters in the flat $w_0 w_a$CDM$+A_L$ model favored by (left) non-CMB, P18, and P18+non-CMB data sets, and (right) non-CMB, P18+lensing, and P18+lensing+non-CMB data sets. The horizontal or vertical dotted lines representing $w_0=-1$, $w_a=0$, $w_0+w_a=-1$ correspond to the values in the standard $\Lambda$CDM model.
}
\label{fig:flat_w0waCDM_Alens_P18_nonCMB23v2_w0wa}
\end{figure*}

We present the $w_0w_a$CDM and $w_0w_a$CDM+$A_L$ dynamical dark energy parameterization results in Tables \ref{tab:results_flat_w0waCDM} --  \ref{tab:consistency_w0waCDM} 
and in Figs.\ \ref{fig:flat_w0waCDM_P18_nonCMB23v2} -- \ref{fig:flat_w0waCDM_Alens_P18_nonCMB23v2_w0wa}.
When studying the consistency of cosmological parameter constraints between P18 and non-CMB data and between P18+lensing and non-CMB data, we use two statistical estimators, $\log_{10}\mathcal{I}$ and $p$ values. These results are presented in Table \ref{tab:consistency_w0waCDM}. The $\Delta \chi^2_{\rm min}$, $\Delta$AIC, and $\Delta$DIC values are provided in Tables \ref{tab:results_flat_w0waCDM} and \ref{tab:results_flat_w0waCDM_AL}. 

Due to the minor role played by dark energy at the redshift of the CMB data, the constraints on $w_0$ and $w_a$ coming from the analysis of P18 and P18+lensing data are weaker than those obtained with non-CMB data. The same is true for the derived parameters $H_0$, $\Omega_m$, and $\sigma_8$. 

As discussed in Ref.\ \refcite{Park:2024jns}, within the context of the $w_0w_a$CDM parameterization, there exists a tension between P18 and non-CMB data cosmological constraints and also between P18+lensing and non-CMB data cosmological constraints. In this work we investigate whether allowing the lensing consistency parameter $A_L$ to vary (see Refs.\ \refcite{deCruzPerez:2022hfr} and \refcite{deCruzPerez:2024abc} for a detailed study of the inclusion of this parameter in other cosmological models) can help alleviate these discrepancies. From the values presented in Table \ref{tab:consistency_w0waCDM}, we see in the P18 vs.\ non-CMB case that while the $w_0w_a$CDM model with $A_L=1$ yields $\log_{10} \mathcal{I} = -0.891$ for the first estimator, the $w_0w_a$CDM+$A_L$ model gives $\log_{10} \mathcal{I} = 0.079$. For the second estimator, when $A_L =1$, we find $\sigma = 2.801$ whereas allowing $A_L$ to vary reduces it to $\sigma=1.896$. Regardless of the estimator considered, the results show that when $A_L$ is allowed to vary, particularly when $A_L>1$, the tension between P18 and non-CMB data cosmological constraints is reduced. This reduction in the differences is milder when CMB lensing data are included in the analysis. In studying the tension between P18+lensing and non-CMB data, we find that for the $w_0w_a$CDM parameterization with $A_L=1$, $\log_{10} \mathcal{I} = -0.787$ and $\sigma = 2.653$, whereas for the $w_0w_a$CDM+$A_L$ parameterization we obtain $\log_{10} \mathcal{I} = -0.030$ and $\sigma = 2.050$. While, as expected, there is a reduction in the level of differences compared to the $A_L=1$ case, we still find a $\sigma >2$ (but $< 3$) tension between the cosmological parameter constraints obtained with P18+lensing data and with non-CMB data. In light of these results, we may conclude that P18, lensing, and non-CMB data can also be jointly analyzed within the context of the dynamical dark energy $w_0w_a$CDM+$A_L$ parameterization, as in the $A_L = 1$ case \cite{Park:2024jns}. Consequently, we largely focus on results from the analysis of the P18+lensing+non-CMB data, the largest data compilation we study. 


We present the $w_0w_a$CDM and $w_0w_a$CDM+$A_L$ dynamical dark energy parameterization results in Tables \ref{tab:results_flat_w0waCDM} --  \ref{tab:consistency_w0waCDM} 
and in Figs.\ \ref{fig:flat_w0waCDM_P18_nonCMB23v2} -- \ref{fig:flat_w0waCDM_Alens_P18_nonCMB23v2_w0wa}.
When studying the consistency of cosmological parameter constraints between P18 and non-CMB data and between P18+lensing and non-CMB data, we use two statistical estimators, $\log_{10}\mathcal{I}$ and $p$ values. These results are presented in Table \ref{tab:consistency_w0waCDM}. The $\Delta \chi^2_{\rm min}$, $\Delta$AIC, and $\Delta$DIC values are provided in Tables \ref{tab:results_flat_w0waCDM} and \ref{tab:results_flat_w0waCDM_AL}. 

\begin{table}[htpb]
\tbl{Consistency check parameter $\log_{10} \mathcal{I}$ and tension parameters $\sigma$ and $p$ for P18 vs.\ non-CMB data sets and P18+lensing vs.\ non-CMB data sets in the flat $w_0 w_a$CDM$(+A_L)$ parameterizations.}
{\begin{tabular}{@{}lcc@{}} \toprule
                                 & \multicolumn{2}{c}{Flat $w_0 w_a$CDM parameterization}  \\[+1mm]
\cline{2-3}\\[-1mm]
   Data                          &  P18 vs non-CMB  & P18+lensing vs non-CMB  \\
\colrule
  $\log_{10} \mathcal{I}$        &   $-0.891$       &  $-0.787$       \\
  $\sigma$                       &   $2.801$        &  $2.653$        \\
  $p$ (\%)                       &   $0.509$        &  $0.798$        \\
\colrule
                                 & \multicolumn{2}{c}{Flat $w_0 w_a$CDM$+A_L$ parameterization}    \\[+1mm]
\cline{2-3}\\[-1mm]
  Data                           &   P18 vs non-CMB & P18+lensing vs non-CMB    \\
\colrule
  $\log_{10} \mathcal{I}$        &   $0.079$        &  $-0.030$        \\
  $\sigma$                       &   $1.896$        &  $2.050$        \\
  $p$ (\%)                       &   $5.800$        &  $4.032$       \\
\botrule
\end{tabular}\label{tab:consistency_w0waCDM}}
\end{table}

From Tables \ref{tab:results_flat_w0waCDM} and \ref{tab:results_flat_w0waCDM_AL} we can compare the results obtained with P18+lensing+non-CMB data for the eight-parameter $w_0w_a$CDM model and the nine-parameter $w_0w_a$CDM+$A_L$ model. For the six primary parameters common to the flat $\Lambda$CDM model, the shifts in the values remain below $1\sigma$, in particular: $\Omega_b h^2$ ($-0.58\sigma$), $\Omega_c h^2$ ($+0.90\sigma$), $100\theta_{\text{MC}}$ ($-0.36\sigma$), $\tau$ ($+0.46\sigma$), $n_s$ ($-0.71\sigma$), and $\ln(10^{10}A_s)$ ($+0.66\sigma$). In regard to the equation of state parameters, while for the $w_0w_a$CDM parameterization we have $w_0=-0.850\pm 0.059$ and $w_a=-0.59^{+0.26}_{-0.22}$ for the $w_0w_a$CDM+$A_L$ parameterization we find $w_0=-0.879\pm 0.060$ and $w_a=-0.39^{+0.26}_{-0.22}$, with the difference between these pair of values being $+0.34\sigma$ and $-0.59\sigma$. Additionally, we provide the value for the combination $w_0+w_a$. For the $w_0w_a$CDM model $w_0+w_a=-1.44^{+0.20}_{-0.17}$ and for the $w_0w_a$CDM+$A_L$ model $w_0+w_a=-1.27^{+0.20}_{-0.17}$, with a tension of $-0.65\sigma$ between the two results. As for the derived parameters, $H_0$, $\Omega_m$, and $\sigma_8$, the differences between the values are $-0.04\sigma$, $+0.36\sigma$, and $1.12\sigma$, respectively. 

We now examine whether the P18+lensing+non-CMB data compilation provides model-independent cosmological parameter constraints. The results for the flat $\Lambda$CDM model, obtained after analysing the P18+lensing+non-CMB data, are presented in the right column of the upper half of Table IV in Ref.\ \refcite{deCruzPerez:2024abc}, while the corresponding results for the flat $w_0w_a$CDM+$A_L$ parameterization can be found in Table \ref{tab:results_flat_w0waCDM_AL}. For the six common primary parameters, the differences are: $-0.35\sigma$ for $\Omega_b h^2$, $+0.54\sigma$ for $\Omega_c h^2$, $-0.15\sigma$ for $100\theta_{\text{MC}}$, $+0.78\sigma$ for $\tau$, $-0.46\sigma$ for $n_s$, and $+0.92\sigma$ for $\ln(10^{10}A_s)$. Although all the differences remain below $1\sigma$, it is noteworthy that when $A_L$ is allowed to vary in the analysis, the shift in the cosmological parameter values, except for the $100\theta_{\text{MC}}$ parameter, are greater than those obtained when comparing the results for the flat $w_0w_a$CDM (with $A_L=1$) model and the flat $\Lambda$CDM model, see discussion in Sec.\ IV of Ref.\ \refcite{Park:2024jns}. For the derived parameters $H_0$, $\Omega_m$, and $\sigma_8$, the differences are at $+0.28\sigma$, $-0.04\sigma$, and $+1.03\sigma$ respectively. 

In order to compare the performance of the flat $\Lambda$CDM model with that of the $w_0w_a$CDM+$A_L$ dynamical dark energy parameterization, it is important to account for the extra parameters in the latter model, which has three additional degrees of freedom $w_0$, $w_a$ and $A_L$. In this work we use the AIC and DIC to properly penalize the presence of these extra parameters (see Sec.\ \ref{sec:Methods} for details on these statistical estimators). According to the information criteria considered here, both $\Delta\textrm{DIC} = -4.37$ and $\Delta\textrm{AIC} = -5.00$ indicate that the $w_0w_a$CDM+$A_L$ parameterization is {\it positively} favored over the standard model. This means that including the $A_L$ parameter in the analysis improves the model's ability to fit these observational data. When performing the same comparison for the $w_0w_a$CDM model with $A_L=1$, we find $\Delta\textrm{DIC} = -2.45$ and $\Delta\textrm{AIC} = -2.25$, which still indicates that this dynamical dark energy parameterization is also {\it positively} favored over the $\Lambda$CDM model, though the magnitudes of $|\Delta\text{DIC}|$ and $|\Delta\text{AIC}|$ are reduced by about $\sim 2$ units. Finally, comparing the performance of the $w_0w_a$CDM model ($A_L=1$) with that of $w_0w_a$CDM+$A_L$ model, we see that the latter has smaller DIC and AIC than the former by $1.92$ and $2.75$, respectively, indicating that the model with a varying $A_L$ is on the verge of being {\it positively} favored. 

As previously stated, in light of the results presented in Table \ref{tab:consistency_w0waCDM}, the variation of the lensing consistency parameter $A_L$ helps reduce the tension between P18 and non-CMB data cosmological constraints and also between P18+lensing and non-CMB data cosmological constraints. This can  be seen more visually by comparing the contour plots displayed in Figs.\ \ref{fig:flat_w0waCDM_P18_nonCMB23v2_w0wa} and \ref{fig:flat_w0waCDM_Alens_P18_nonCMB23v2_w0wa}. In the case with $A_L=1$, the contours in the $w_0-w_a$ plane, obtained with P18 data and with non-CMB data, do not overlap at $\sim 2\sigma$ level. However, in the parameterization with a varying $A_L$, an overlap occurs even at the $1\sigma$ level. The same is true for the comparison between P18+lensing data and non-CMB data contours. 

From the P18+lensing+non-CMB data set in the flat $w_0w_a$CDM$+A_L$ parameterization we get $H_0=67.84\pm 0.64$ km s$^{-1}$ Mpc$^{-1}$, which agrees with the median statistics result $H_0=68\pm 2.8$ km km s$^{-1}$ Mpc$^{-1}$ \cite{Chen:2011ab,Gottetal2001,Calabreseetal2012}, as well as with some local measurements including the flat $\Lambda$CDM model value of Ref.\ \refcite{Cao:2023eja} $H_0=69.25\pm 2.4$ km s$^{-1}$ Mpc$^{-1}$ from a joint analysis of $H(z)$, BAO, Pantheon+ SNIa, quasar angular size, reverberation-measured \mii\ and \civ\ quasar, and 118 Amati correlation gamma-ray burst data, and the local $H_0=69.03\pm 1.75$ km s$^{-1}$ Mpc$^{-1}$ from JWST TRGB+JAGB and SNIa data \cite{Freedman:2024eph}, but is in tension with the local $H_0=73.04\pm 1.04$ km s$^{-1}$ Mpc$^{-1}$ measured using Cepheids and SNIa data \cite{Riess:2021jrx}, also see Ref.\ \refcite{Chen:2024gnu}. And the flat $w_0w_a$CDM$+A_L$ parameterization P18+lensing+non-CMB data value $\Omega_m = 0.3062\pm 0.0064$ also agrees well with the flat $\Lambda$CDM model value of $\Omega_m = 0.313\pm 0.012$ of Ref.\ \refcite{Cao:2023eja} (for the data set listed above used to determine $H_0$).

Comparing the blue $w_0$--$w_a$ likelihood contours of the flat $w_0 w_a$CDM parameterization for the P18+lensing+non-CMB data, shown in the right panel of Fig.\ \ref{fig:flat_w0waCDM_P18_nonCMB23v2_w0wa}, to the corresponding flat $w_0 w_a$CDM$+A_L$ parameterization blue contours in the right panel of Fig.\ \ref{fig:flat_w0waCDM_Alens_P18_nonCMB23v2_w0wa}, we see that the upper left vertex of the $w_0 w_a$CDM parameterization $2\sigma$ blue contour almost touches the flat $\Lambda$CDM model point of $w_0 = -1$ and $w_a = 0$ \cite{Park:2024jns}, while the flat $\Lambda$CDM model point is just ouside the $1\sigma$ contour in the $w_0 w_a$CDM$+A_L$ parameterization case. As noted in Ref.\ \refcite{Park:2024jns} our $w_0 w_a$CDM parameterization P18+lensing+non-CMB data result is consistent with, but more restrictive than the corresponding DESI+CMB+PantheonPlus result of Ref.\ \refcite{DESI:2024mwx}, reflecting the greater constraining power of our data compilation. We also showed in Ref.\ \refcite{Park:2024jns} that even when we exclude Pantheon+ SNIa data from our non-CMB data compilation the flat $\Lambda$CDM model is still $\sim 2\sigma$ away from the best-fit dynamical dark energy model, so the $\sim2\sigma$ support we find for dark energy dynamics in this parameterization is neither caused by DESI BAO data (which we have not used), nor caused by Pantheon+ SNIa data.

In our analyses here we find that in the $w_0 w_a$CDM$+A_L$ parameterization for the P18+ lensing+non-CMB data we get $A_L = 1.078^{+0.036}_{-0.040}$ meaning that the option $A_L > 1$ is preferred at $1.95\sigma$. Additionally, this parameterization simultaneously favors the possibility of having a time-evolving dark energy component over a cosmological constant by approximately $\sim 1\sigma$. This can be seen either, as noted above, by looking at the $w_0-w_a$ sub-panel in the right panel of Fig.\ \ref{fig:flat_w0waCDM_Alens_P18_nonCMB23v2_w0wa} or by looking at the asymptotic value of the equation of state parameter at high redshift $w_0+w_a$, whose value in the standard flat $\Lambda$CDM model is $w_0+w_a=-1$, but instead in the flat $w_0w_a$CDM+$A_L$ parameterization takes the value $w_0+w_a = -1.27^{+0.20}_{-0.17}$ representing a deviation from $w_0+w_a=-1$ of $1.35\sigma$, less than the $2.2\sigma$ deviation we find from $w_0+w_a = -1.44^{+0.20}_{-0.17}$ in the $w_0w_a$CDM ($A_L=1$) parameterization. 

We showed in Ref.\ \refcite{Park:2024jns} for the $w_0w_a$CDM parameterization that P18 data likelihood contours clearly favor a phantom dark energy equation of state, high values of $H_0$, and low values of $\Omega_m$. Including non-CMB data in the mix breaks the degeneracy between parameters, pushing $H_0$ down, $\Omega_m$ up, and decreasing the evidence in favor of phantom dark energy. We find the same is true for the $w_0w_a$CDM$+A_L$ parameterization, but now for the P18+lensing+non-CMB data set there is only $\sim 1\sigma$ evidence in favor of dynamical dark energy but $\sim 2 \sigma$ evidence in favor of $A_L > 1$, i.e., for more weak CMB lensing than is predicted by the best-fit cosmological model. So while the $\sim2\sigma$ support we find for dark energy dynamics in the $w_0w_a$CDM parameterization is neither caused by DESI BAO data nor caused by Pantheon+ SNIa data, our results suggest that the excess weak CMB lensing smoothing seen in Planck PR3 data might be contributing to the evidence for dark energy dynamics seen in the $w_0w_a$CDM parameterization. 

It is important to note that consideration of the new Planck data release (PR4) \cite{Tristram:2023haj}, which is the updated version of the Planck 2018 TT,TE,EE+lowE data (denoted as PR3, for comparison, here) employed in this work, can induce some changes in the values of the cosmological parameters. The authors of Ref.\ \refcite{Tristram:2023haj} find from PR4 data in the flat $\Lambda$CDM+$A_L$ model a lensing consistency parameter $A_L = 1.039\pm 0.052$ (with $0.75\sigma$ evidence in favor of $A_L>1$), whereas when PR4 data are considered together with updated weak CMB lensing data $A_L = 1.037\pm 0.037$ (with $1\sigma$ evidence in favor of $A_L>1$). These values are to be compared with $A_L = 1.181\pm 0.067$, obtained from PR3 and showing a preference for $A_L>1$ at $2.7\sigma$, and with $A_L = 1.073\pm 0.041$ obtained from PR3+lensing and $1.78\sigma$ away from $A_L=1$, respectively \cite{deCruzPerez:2022hfr}. According to these results the excess amount of weak gravitational lensing present in the PR4 version of the CMB power spectra is smaller than in the corresponding version of the PR3 and consequently the evidence in favor of $A_L>1$ is reduced. Another important point is that, while with PR3 data the joint use of PR3 and lensing data reduces the evidence in favor of $A_L>1$ with respect to the analysis with PR3 data alone, in the corresponding case with the PR4 versions of the likelihoods it is the other way around. As noted, part of the evidence in favor of a dynamical dark energy component seems to come from the excess of lensing observed in the Planck CMB anisotropy spectra. Therefore, if PR4 data reduce the effect of the lensing anomaly we expect that the evidence for a time-evolving dark energy density will be reduced when PR4 (instead of PR3) data are analyzed with non-CMB data.

\section{Conclusion}
\label{sec:Conclusion}

We have tested the flat dynamical dark energy $w_0w_a$CDM+$A_L$ parameterization, with a varying lensing consistency parameter $A_L$, that is characterized by a dynamical dark energy fluid equation of state parameter $w(z) = w_0 + w_{a}z/(1+z)$, with different data set combinations including CMB and non-CMB data. For the most complete data set, P18+lensing+non-CMB, we find $w_0=-0.879\pm 0.060$ and $w_a=-0.39^{+0.26}_{-0.22}$, with the high-redshift asymptotic limit $w_0+w_a=-1.27^{+0.20}_{-0.17}$, and $A_L=1.078^{+0.036}_{-0.040}$. Therefore time evolution of the dark energy fluid is favored over a $\Lambda$ at $\sim 1\sigma$ and a preference for $A_L>1$ is found with a significance of $\sim 2\sigma$. When its performance is compared with that of the flat $\Lambda$CDM model and with that of the flat $w_0w_a$CDM parameterization with $A_L=1$, the $w_0w_a$CDM+$A_L$ parameterization turns out to be {\it positively} favored by these data. Moreover, the variation of the phenomenological lensing consistency parameter $A_L$ helps in better reconciling the P18 and non-CMB and the P18+lensing and non-CMB cosmological parameter constraints with respect to the case with $A_L=1$. 

While the evidence for dark energy dynamics in the $w_0w_a$CDM parameterization \cite{DESI:2024mwx, Park:2024jns} does not depend on DESI BAO data \cite{Park:2024jns}, nor on the use of Pantheon+ SNIa data \cite{Park:2024jns}, our results here suggest that it at least partially depends on the excess smoothing seen in some of the Planck CMB anisotropy multipoles. 

While these results are interesting, they are not that statistically significant. Additionally $w_0w_a$CDM+$A_L$ is not a physically consistent dynamical dark energy model but rather just a parameterization. However, as we have shown here, allowing for a varying lensing consistency parameter does alleviate some tensions between cosmological parameter constraints obtained with different data sets that are present when $A_L = 1$, and it also provides a better fit to these data than is provided by the standard flat $\Lambda$CDM model as well as the $w_0w_a$CDM parameterization. Our results therefore motivate a more careful analysis, with better and more data, of the $w_0w_a$CDM$(+A_L)$ parameterizations.

\section*{Acknowledgments}

J.d.C.P.'s research was financially supported by the project "Plan Complementario de I+D+i en el \'area de Astrof{\'\i}sica" funded by the European Union within the framework of the Recovery, Transformation and Resilience Plan - NextGenerationEU and by the Regional Government of Andaluc{\'i}a (Reference AST22\_00001). C.-G.P.\ was supported by a National Research Foundation of Korea (NRF) grant funded by the Korea government (MSIT) No.\ RS-2023-00246367.

\bibliographystyle{ws-ijmpd}
\bibliography{w0waCDM_Alens}

\end{document}